# WELFARE WITHOUT TAXATION

## Autonomous production revenues for Universal Basic Income

Eleanor 'Nell' Watson FCBS FIAP FRSA FRSS FLS CITP     and Doreen Bianca

**Keywords;** Universal Basic Income, Autonomous production, Artificial Intelligence

## I. Introduction

In the face of shifting means of production from manual human labor to labor automation, one solution that stands out is the advancement of a Universal Basic Income, UBI to every citizen from the government with no strings attached. The proposal, however, has encountered sharp criticism from different quarters questioning the morality behind sourcing of funds, largely through taxation, to uphold an institution designed to provide social support. Others also perceive the idea as a form of socialism, or a capitalist road to communism (1). The current discussion, however, seeks to demonstrate that the provision of such stipend can occur through the utilization of revenues realized from production driven by Artificial Intelligence (AI), and to a small extent, philanthropic contributions from the top 1 percent of the population.

## II. Background of UBI debates

Disagreements regarding modern, more pragmatic means of handling welfare programs tend to focus on three pertinent issues. First, the existence of two opposing ideological perceptions that on one side demand for improved welfare spending accompanied by higher taxes, while the other side is championing for decreased welfare spending, and subsequently lower taxes (2). Second, an economist-driven welfare reform process that is characteristically promising in theory, but which lacks proper design of the transition process that makes it politically feasible (3). And finally, the fact that debates on welfare reforms remain typically narrow, hardly focusing on reforms capable of delivering comprehensive rules to govern, or better still, reconcile the welfare and tax systems (4).

However, the pertinent question given the current technological advancement, is whether debates on welfare should continue focusing on the taxation aspect. In the wake of technological advancements that have seen robots powered by AI take up jobs that, during the industrial revolution and post-industrial revolution eras served a greater fraction of the populace, there is need for a means of tackling the ever-increasing income disparity within the populace. Within the context of this consequential disruption, it is increasingly becoming evident that individuals who cannot or are unable to mount successful competition against the advanced and talent-driven economies currently observed will require welfare assistance in the form of Universal Basic Income, UBI. Contrary to traditional, and largely the prevailing belief that intervention through UBI is dependent on taxation, specifically skewed towards increased taxation that will affect the very beneficiaries of UBI, the focus ought to shift to the revenues obtained from AI driven technologies.

## III. Autonomous production and revenue generation

Leveraging technology for increased production capacity at minimal cost is the new norm. The blockchain technology, for instance, led to the emergence of smart transactions, defined by contracts that automatically process payment once the transaction is complete (5)(6). Taken further, embedding AI into such smart contracts has resulted in the emergence of Distributed Autonomous Organizations, DAOs. Such organizations have proven capable of running themselves, specifically by being able to conduct and complete trading activities, as well as hiring humans to perform certain tasks. The obvious impact here is that having successfully replaced humans in the working class labor and middle-class clerical capacities, right on its sight with the emergence of the iCEO are the middle-class management and the C-suite. The presence of these capabilities means that the world is at a point where it is practically possible to create wealth voluntarily and within the principles of the free market.

The resulting wealth, however, does have proprietors, who are either the owners of the business or the shareholders. In a capitalist economy, it would be impossible challenging an individual's right to ownership of such colossal wealth. However, in the face of the existence of renowned philanthropists such as Bill and Melinda Gates Foundation, Warren Buffet, George Soros, Mark Zuckerberg, among others (7), the government can design a comprehensive mechanism through which to channel such resources towards the fulfillment of UBI. Noteworthy here is that such philanthropy tends to target causes and populations in developing and underdeveloped countries. UBI, on the contrary, will specifically target the nation's citizens and will be practically impossible to apply beyond the borders of the nation. However, this does not imply the impossibility of continued support for the development of developing nations, either in terms of philanthropic disbursements or financial and material aid from one nation-state government to another as is currently the case.

The sources of funds for UBI so far described are on a purely voluntary basis. Rather surprisingly, the software technology realm currently boasts the existence of a Free and Open-Source Software (FOSS) movement whose objective is to enable individuals from across the world access, use, study, copy or even modify in any manner suitable to the user, a software, and shares the source code of the software rather openly to encourage the voluntary improvement of the software. While pointing at the benefits of FOSS, which include, though not limited to decreased software costs, improved security, improved privacy since it gives users more control of their hardware, Soderberg argues that the movement rather "opens up the practice of intervening in computer technology to a non-denumerable mass of people," (8). The movement encourages user-centered production, which, from an analytical perspective, has better chances of outperforming the conventional capitalist markets in the provision of social needs. The case of FOSS is much that of the capitalist market's failure to satisfy human needs, which in turn motivates individuals to abandon the conventional market relations.

A deeper examination of the character and nature of those involved in the contributions so far realized by FOSS shows that a greater fraction, if not the entire team, comprise of random nerds voluntarily producing things for free, in return for a small proportion of status within a confined society, and whose desire is to make the world a better place with their ingenuity. Free and open source businesses running on AI and capable of trade, provisioning of microservices rather autonomously will ride on a similar premise. Such businesses can produce real value for real people. In addition, they will be capable of running charities for free. For instance, such mechanism would see the emergence of mutual insurance funds that run themselves. As such, any area currently characterized by low profit margins and attracting hardly any interest from human-led organizations will be given a new lease of life. The resources and revenues so realized can go into taking care of UBI payments.

Furthermore, DAOs will also be capable of producing dividends for real shareholders. All it would require is some form of capability and the FOSS mindset to bootstrap an open business within a short period, resulting in common people obtaining real improvements in their lifestyles. Interestingly, this mechanism requires no form of taxation or coercion, with any successful creation resulting in huge gains for the one behind the idea. While such firms already exist, such as Newman's Own and co-operatives, integrating the power of AI into such structures will enable the expansion of such ventures to levels currently unimaginable. Note that DAOs will not eliminate traditional companies in their entirety, which implies that there will be a mixed market scenario that prides both for-profit and non-profit ventures.

That we are at the early days of automation of thinking is irrefutable. Having retired a greater fraction of bureaucratic and clerical duties that were the preserve of human beings, computer Algorithms will soon begin taking over the more creative tasks. Terrifying as it may be, we are in an era where Netflix has algorithms capable of commissioning which movies an individual should watch and a number of news outlets have algorithms used to generate content on crime, games, and other natural events, which points to the invasion of the most rote tasks in journalism (9). Composition and production of fine art can also occur using computer algorithms. Given the mentioned trajectory of technological advancement, sooner it will be of no use nurturing human creativity. Yet, demand for social services will remain, and advancing UBI to citizens from the proceeds of these autonomous processes would be the sole available option. Such revenues will not come from tax, nor will they come from philanthropic contributions, they will be the result of AI-driven production.

A study conducted by the Roosevelt Institute on the feasibility of UBI indicates that advancing a given amount of stipend to mem-bers of the society by growing the federal debt, i.e., without raising taxes, would result in a substantial growth of the economy. According to its finding, advancing, say $12000 per year per adult would result in a permanent economic growth equivalent to between 12.56 and 13.10 percent, or about 2.5 trillion dollars by 2025 (10). In addition, such a policy would result in a 2 percent increase in the number of Americans with jobs (10). Given the undesirability of federal debt expansion by a good number of contemporary economists, the current study aims to show that the results would be rosier if such UBI funding does not come from debt expansion, but rather from actual revenue realized from AI-driven autonomous production.

## IV. Conclusion

In recent years, discussion of the potential implementation of UBI has pervaded every aspect of social and economic policies, especially given that many stakeholders felt that it would be foolish for any struggling government infrastructure to assume such colossal responsibility. Often, the impediments have been largely the sources of funds since majority analysts have always cast their eyes on taxation, and to some extent, government borrowing. However, with the emergence of a new model of wealth creation, namely through autonomous production mechanism driven by DAOs and AI, the long pervading conundrum is no more. Technological developments continue to render many people jobless by eliminating direct human intervention in different fields of human interest. Given that production and wealth creation continues to take place, it becomes logical that governments should draw comprehensive plans to initiate and sustain UBI for the citizens funded by revenues from autonomous production, thus completely eliminating the need for increased taxation to support welfare.